\documentclass[default]{aastex62}

\usepackage{graphicx}	% Including figure files
\usepackage{amsmath}	% Advanced maths commands
\usepackage{amssymb}	% Extra maths symbols
\usepackage{array}
\usepackage{isotope}
\usepackage{mathtools}

\usepackage[flushleft]{threeparttable}
\usepackage{hyperref}
\hypersetup{
    colorlinks=true,
    linkcolor=blue,
    filecolor=magenta,      
    urlcolor=blue,
}
\usepackage{mathtools}
\DeclareUnicodeCharacter{2212}{-}
\graphicspath{{./}{figures/}}

\shorttitle{HC(O)SH and C$_2$H$_5$SH in the ISM}
\shortauthors{Rodr\'{\i}guez-Almeida et al.}
\begin{document}

\title{Thiols in the ISM: first detection of HC(O)SH and confirmation of C$_2$H$_5$SH}

\correspondingauthor{Lucas F. Rodr\'{\i}guez-Almeida}
\email{lrodriguez@cab.inta-csic.es}

\author{Lucas F. Rodr\'{\i}guez-Almeida}
\affil{Centro de Astrobiolog\'{\i}a (CSIC-INTA), Ctra Ajalvir km 4, 28850, Torrej\'{o}n de Ardoz, Madrid, Spain}

\author{Izaskun Jim\'{e}nez-Serra}
\affiliation{Centro de Astrobiolog\'{\i}a (CSIC-INTA), Ctra Ajalvir km 4, 28850, Torrej\'{o}n de Ardoz, Madrid, Spain}

\author{V\'{\i}ctor M. Rivilla}
\affiliation{Centro de Astrobiolog\'{\i}a (CSIC-INTA), Ctra Ajalvir km 4, 28850, Torrej\'{o}n de Ardoz, Madrid, Spain}
\affiliation{INAF-Osservatorio Astrofisico di Arcetri, Largo Enrico Fermi 5, 50125, Florence, Italy}

\author{Jes\'us Mart\'{\i}n-Pintado}
\affiliation{Centro de Astrobiolog\'{\i}a (CSIC-INTA), Ctra Ajalvir km 4, 28850, Torrej\'{o}n de Ardoz, Madrid, Spain}

\author{Shaoshan Zeng}
\affiliation{Star and Planet Formation Laboratory, Cluster for Pioneering Research, RIKEN,2-1 Hirosawa, Wako, Saitama, 351-0198, Japan}

\author{Bel\'en Tercero}
\affiliation{Observatorio de Yebes (IGN), Cerro de la Palera s/n, 19141, Guadalajara, Spain}

\author{Pablo de Vicente}
\affiliation{Observatorio de Yebes (IGN), Cerro de la Palera s/n, 19141, Guadalajara, Spain}

\author{Laura Colzi}
\affiliation{Centro de Astrobiolog\'{\i}a (CSIC-INTA), Ctra Ajalvir km 4, 28850, Torrej\'{o}n de Ardoz, Madrid, Spain}
\affiliation{INAF-Osservatorio Astrofisico di Arcetri, Largo Enrico Fermi 5, 50125, Florence, Italy}

\author{Fernando Rico-Villas}
\affiliation{Centro de Astrobiolog\'{\i}a (CSIC-INTA), Ctra Ajalvir km 4, 28850, Torrej\'{o}n de Ardoz, Madrid, Spain}

\author{Sergio Mart\'{\i}n}
\affiliation{Eureopean Southern Observatory, Alonso de C\'{o}rdova 3107, Vitacura 763 0355, Santiago, Chile}
\affiliation{Joint ALMA Observatory, Alonso de C\'{o}rdova 3107, Vitacura 763 0355, Santiago, Chile}

\author{Miguel A. Requena-Torres}
\affiliation{University of Maryland, College Park, ND 20742-2421, USA}
\affiliation{Department of Physics, Astronomy and Geosciences, Towson University, MD 21252, USA}

\begin{abstract}

The chemical compounds carrying the thiol group (-SH) have been considered essential in recent prebiotic studies regarding the polymerization of amino acids. We have searched for this kind of compounds toward the Galactic Centre quiescent cloud G+0.693-0.027. We report the first detection in the interstellar space of the trans-isomer of monothioformic acid (t-HC(O)SH) with an abundance of $\sim\,$1$\,\times\,$10$^{-10}$. Additionally, we provide a solid confirmation of the gauche isomer of ethyl mercaptan (g-C$_2$H$_5$SH) with an abundance of $\sim\,$3$\,\times\,$10$^{-10}$, and we also detect methyl mercaptan (CH$_3$SH) with an abundance of $\sim\,$5$\,\times\,$10$^{-9}$. Abundance ratios were calculated for the three SH-bearing species and their OH-analogues, revealing similar trends between alcohols and thiols with increasing complexity. Possible chemical routes for the interstellar synthesis of t-HC(O)SH, CH$_3$SH and C$_2$H$_5$SH are discussed, as well as the relevance of these compounds in the synthesis of prebiotic proteins in the primitive Earth.

\end{abstract}



\section{Introduction} 
\label{sec:intro}

Among different theories of origin of life, one recurrent conundrum is the abiotic polymerization of amino acids since it requires ribosomes, macromolecular machines containing ribonucleic acid (RNA) and proteins. How could the first proteins form if they were needed to synthesize others? Following the ideas of %\citet{liu1997}
\citet[][]{foden2020prebiotic}, a possible solution involves a thiol-based scenario in which SH-bearing molecules, together with the family of thioacids (R-C(O)SH) and thioesters (R-S-R'), have important properties as energy carriers and catalysts \citep[][]{chandru2016,leman2017}. Although these types of compounds could be created in-situ by an H$_2$S-mediated chemistry under prebiotically plausible conditions on early Earth \citep[][]{shalayel2020}, they also could have been delivered exogenously. Hence, observations of thiol-based molecules in space could shed some light on the availability of such compounds on a primitive Earth, and on their role in the prebiotic synthesis of proteins.

%son 20 S-bearing sin contar con C2H5SH (dubious) y C5S (same)
More than 220 molecules have been detected in the interstellar medium (ISM) and circumstellar shells\footnote{ \url{https://cdms.astro.uni-koeln.de/classic/molecules}} to date. However, sulfur-containing species only account for 20 of them. Furthermore, while molecules detected carrying C, H or N range from 2 up to 13 atoms, the vast majority of S-bearing molecules have, at most, 4 atoms \citep[such as H$_2$CS;][]{sinclair1973}. This could be due to the relatively low cosmic abundance of atomic sulfur \citep[$\approx\,$10$^{-5}$ with respect to H$_2$, that is, more than 10 times lower than C or O;][]{asplund2009} together with its ability to have many different oxidation states and allotropes when compared to the more abundant elements$\,$\citep[][]{jm2011,shingledecker2020sulphur} and its capacity of depleting fast in dense molecular clouds$\,$\citep[][]{laas2019}.

As a consequence, a very few S-bearing molecules containing more than 4 atoms have firmly been detected in the ISM so far. One example is methyl mercaptan (hereafter CH$_3$SH), which has been detected in several environments, such as pre-stellar cores$\,$\citep[][]{gibb2000chemistry}, massive star-forming regions like Sagittarius B2$\,$\citep[][]{linke1979,muller2016}, and Solar-like protostars$\,$\citep[][]{majumdar2016}. The other one is ethyl mercaptan (hereafter C$_2$H$_5$SH) which was tentatively detected toward Orion KL$\,$\citep{kolesnikova2014}. Other searches for complex S-bearing molecules were unsuccessful, such as CH$_3$CHS \citep[thioacetaldehyde;][]{margules2020}, NH$_2$CHS \citep[thioformamide;][]{motiyenko2020} and CH$_3$SC(O)H \citep[S-methyl thioformate;][]{jabri2020}.

%and C$_2$H$_5$SH, which has been reported toward Orion KL.

%Despite the intense search of sulfur-bearing complex organic molecules (COMs.) in interstellar space in recent years \citep[see, for example, the tentative detections of CH$_3$CH$_2$SCH$_3$, NH$_2$CHS or NH$_2$CSCH$_3$;][]{cabezas2020,maris2019,motiyenko2020}, only two have firmly been detected to date: CH$_3$SH, toward many different environments such as However, sulfur-bearing species such as thiol acids, have not been reported in the interstellar medium (ISM) so far.

%el paper de Vastel et. al en L1544 tiene upper limit para ch3sh
In this Letter, we report the first detection in the ISM of the trans isomer of monothioformic acid (hereafter HC(O)SH), the simplest thiol acid. We also report a solid confirmation of the gauche isomer of C$_2$H$_5$SH, together with the detection of CH$_3$SH. These molecules are found toward the quiescent Giant Molecular Cloud G+0.693-0.027 located in the Galactic Center (hereafter G+0.693).
%This source shows relatively high gas kinetic temperatures \citep[between 50 K and 140 K;][]{zeng2018complex} with low H$_2$ densities (\sim10$^4$ cm$^{-3}$) which leads to both low dust temperatures and the sub-thermal excitation of molecules within the region \citep[$\leq$30 K; ][]{rodriguez2000}. 
This source shows a very rich chemistry with up to 40 different complex organic molecules (COMs\footnote{These are usually referred as carbon-based molecules that have 6 or more atoms \citep[][]{herbst2009}.}) detected \citep[see, for example:][]{requena2008galactic,zeng2018complex,rivilla2018,rivilla2019_cyanomehtanimine,rivilla2020prebiotic,ijimenez2020}. Studies suggest that this cloud could be undergoing a cloud-cloud collision \citep[][]{zeng2020}, which induces large-scale shocks that sputter dust grains and that enhance the gas-phase abundance of molecules by several orders of magnitude \citep[][]{requenatorres2006}.

% IN A LETTER, IT IS NOT NECESSARY TO INCLUDE THIS TYPE OF PARAGRAPH HERE. BUT WE KEEP IT JUST IN CASE WE DON'T SUBMIT THE PAPER AS A LETTER

%This paper is organized as follows:
%in Sect. \ref{sec:observations} we have detailed the information about the observations, in Sect. \ref{sec:analysis} we have explained the analysis carried out for the new detection of monothioformic acid and related species, in Sect. \ref{sec:discussion} we have done a discussion about the relative abundances present in the source and the chemical pathways that may lead to the formation of these molecules and in Sect. \ref{sec:conclusions} we detail our conclusions. 

\section{Observations} \label{sec:observations}

%(Puesto que solo mostramos en las tablas de las transiciones detectadas, datos hasta 150 GHz aproximadamente, en esta tabla incluiria solo los datos hasta 172 GHz. Eso nos permitira solicitar mas tiempo en el 30m en el futuro.)
We have used a spectral line survey toward G+0.693 covering several windows between 32$\,$GHz to 172$\,$GHz with an average resolution of 1.5$\,$km$\,$s$^{-1}$, although the final spectral resolution employed in the figures has been smoothed up to 3$\,$km$\,$s$^{-1}$. We stress that this is for a proper line visualization and does not alter in any form the analysis done.

For the observations, we used both the IRAM 30m telescope located at Pico Veleta (Granada, Spain) and the Yebes 40m telescope\footnote{Yebes Observatory is operated by the Spanish Geographic Institute (IGN, Ministerio de Transportes, Movilidad y Agencia Urbana)}
%(project code: 20A008)
(Guadalajara, Spain). The equatorial coordinates of the molecular cloud G+0.693 are $\alpha$(J2000.0)$\,$=$\,$17$^h$47$^m$22$^s$, and $\delta$(J2000.0)$\,$=$\,$−28$^{\circ}$21$'$27$''$. The position switching mode was used in all the observations with the off position located at (-885$''$, 290$''$) from the source.  The line intensity of the spectra was measured in units of $\mathrm{T_A^{\ast}}$ as the molecular emission toward G+0.693 is extended over the beam \citep[][]{requenatorres2006,martin2008,rivilla2018}. In all the observations, each frequency setup was repeated shifting the central frequency by 20-100 MHz in order to identify spurious lines or contamination from the image band. 
 
The IRAM 30m observations were performed during three different sessions in 2019: 10-16th of April, 13-19th of August and 11-15th of December. The dual polarization receiver EMIR was used connected to the fast Fourier transform spectrometers (FFTS), which provided a channel width of 200 kHz in the 3 and 2$\,$mm radio windows. The observations with the Yebes 40m radiotelescope were carried out in February 2020: from the 3rd to 9th and from the 15th  to 22th. In this case, the Nanocosmos Q-band (7$\,$mm) HEMT receiver was used which enables ultra broad-band observations in two linear polarizations$\,$\citep[][]{tercero2020}. The receiver was connected to 16 FFTS providing a channel width of 38 kHz and an instantaneous bandwidth of 18.5 GHz per polarization, covering the frequency range between 31.3 GHz and 50.6 GHz. 

%In the analysis, .
%In Table$\;$\ref{tab:resolutions} we list the frequency windows of the IRAM 30m and Yebes 40m observations, with the final velocity resolution that we have used in this work. 

\section{Analysis and results}\label{sec:analysis}

%In the following, we describe the results from the different molecules with the -SH group detected within our dataset. 

%\subsection{Isomers of HC(O)SH and C$_2$H$_5$SH}

We used the software \textsc{madcuba}\footnote{MAdrid Data CUBe Analysis is a software developed at the Center of Astrobiology in Madrid: \url{https://cab.inta-csic.es/madcuba/Portada.html}}$\,$\citep{madcuba} to perform the data analysis and line identification
%\textsc{madcuba}\footnote{MAdrid Data CUBe Analysis is a software developed at the Center of Astrobiology in Madrid: \url{https://cab.inta-csic.es/madcuba/Portada.html}}$\,$\citep{madcuba} was employed. 
%\textsc{madcuba} uses the spectroscopic data entries from different molecular catalogs %the Jet Propulsion Laboratory, JPL\footnote{\url{https://spec.jpl.nasa.gov/}} \citep{JPL}, and Cologne Database for Molecular Spectroscopy, CDMS\footnote{\url{https://cdms.astro.uni-koeln.de/classic/}} \citep{CDMS}, 
%and performs the simultaneous identification of all rotational transitions for a particular molecular species. 
The Spectral Line Identification and Modelling (SLIM) tool of \textsc{madcuba} uses the spectroscopic data entries from different molecular catalogs, and generates a synthetic spectra based under the assumption of Local Thermodynamic Equilibrium (LTE) conditions, and considering line opacity effects. 
%The \textsc{madcuba-autofit} tool calculates the set of physical parameters that best fit all identified molecular transitions using the Levenberg-Marquardt algorithm$\,$\citep{madcuba} comparing both synthetic and observed spectra. 
The fitted parameters used to reproduce the molecular emission are: column density ($N$), excitation temperature ({\it $T_{\mathrm{ex}}$}), local standard of rest velocity ({\it v$_\mathrm{LSR}$}) and full width at half maximum ({\it FWHM}).

Both HC(O)SH and C$_2$H$_5$SH have two rotamers, associated with the rotation of their respective C-S bond. 
HC(O)SH has one cis (c-HC(O)SH) and one trans (t-HC(O)SH) rotamer, being the former  $\sim\,$330$\,$K higher in energy \citep[][]{hocking1976rotational}. C$_2$H$_5$SH has two degenerated $\pm\,$gauche isomers (g-C$_2$H$_5$SH) and one antiperiplanar.

For our analysis we have used the following spectroscopic entries from the the CDMS\footnote{Cologne Database for Molecular Spectroscopy$\,$\citep[][]{CDMS}. \url{https://cdms.astro.uni-koeln.de/classic/}} catalog:
entries 062515/062516 for trans/cis isomers of HC(O)SH \citep[][]{hocking1976rotational}, 062523/062524 for gauche/anti isomers of C$_2$H$_5$SH \citep[][]{kolesnikova2014,muller2016}, and 048510 for CH$_3$SH \citep[][]{xu2012,zakharenko2019}.

\subsection{Detection of t-HC(O)SH}

%Figure 1: transitions of t-HC(O)SH
\begin{figure*}
    \centering
    \includegraphics[width=1\textwidth]{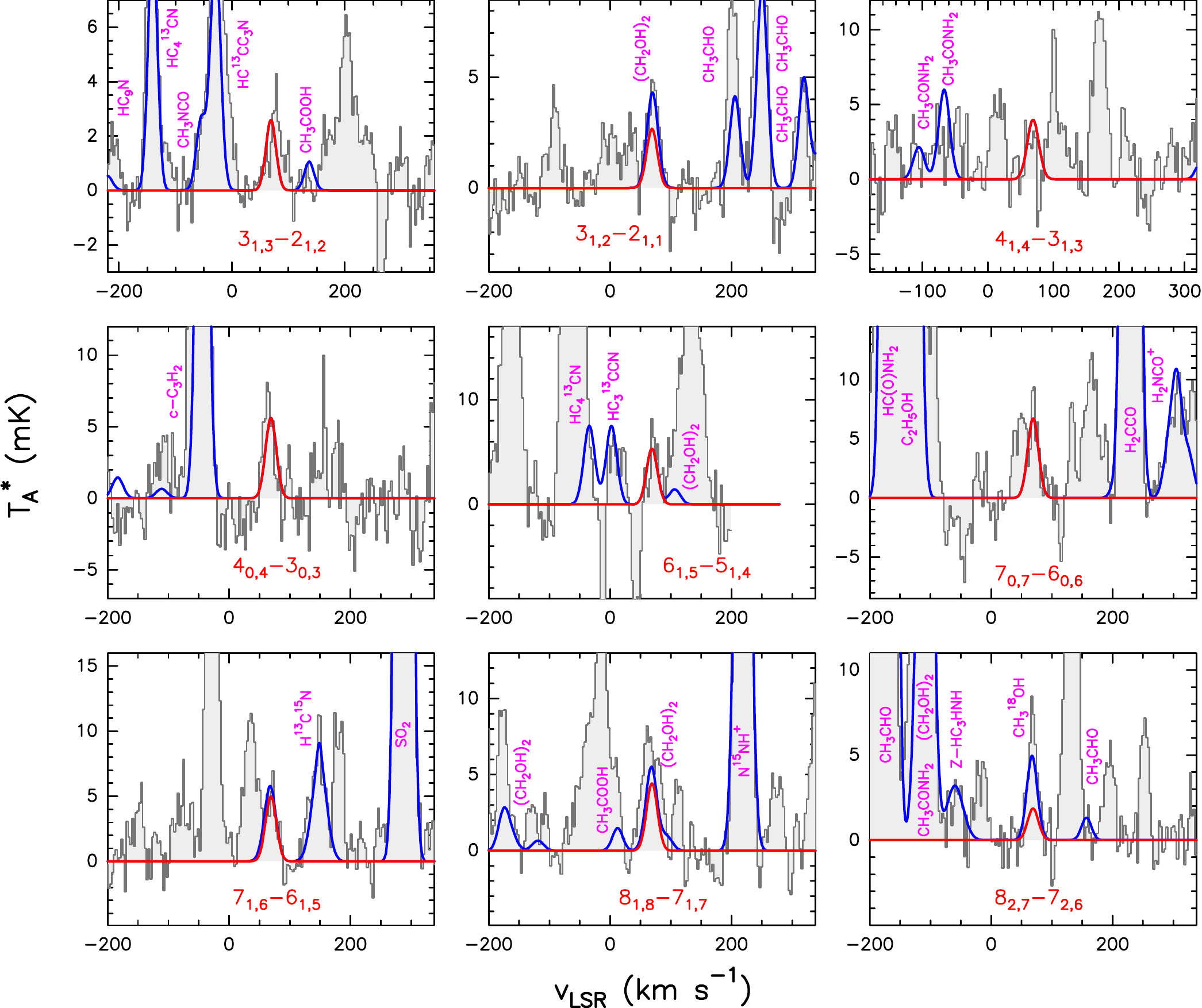}
    \caption{Cleanest and brightest lines of t-HC(O)SH detected toward G+0.693 labeled with their corresponding quantum numbers in red. The red line shows the best LTE fit to the observed spectra (represented by the black lines). The data has been smoothed up to 3 km s$^{-1}$ for an optimal line visualization. The blue lines show the spectra including the emission of all the molecules searched toward G+0.693. Note that these lines are tagged with their corresponding molecular compound in pink.}
    \label{fig:monothioformic}
\end{figure*}

%Table 1:t-HC(O)SH lines
\begin{deluxetable*}{ccccccccl}[htp!]
\tablecaption{Lines of t-HC(O)SH detected toward G+0.693 with their corresponding quantum numbers (QNs), logarithm of the Einstein coefficients ($\mathrm{\log A_{ul}}$) , degeneracy ($\mathrm{g_u}$) and energy ($\mathrm{E_u}$) of the upper state. The integrated signal ($\int T_A^* d\nu$) and root mean square (rms) noise level are also provided and used to calculate the signal to noise ratio (SNR) of the detections.\label{tab:thioformicacid}}
\tablecolumns{9}
\tablewidth{0pt}
\tablehead{
\colhead{Rest frequency} & \colhead{QNs\tablenotemark{a}} & \colhead{g$_u$} & \colhead{E$_u$} & \colhead{$\mathrm{\log A_{ul}}$} & \colhead{rms}& \colhead{$\int T_A^* d\nu$\tablenotemark{b}} & \colhead{SNR\tablenotemark{c}} & \colhead{Comments} \\
\colhead{(MHz)} & \colhead{} & \colhead{} & \colhead{(K)} & \colhead{(s$^{-1}$)} & \colhead{(mK)} & \colhead{($\mathrm{mK\,km\,s^{-1}}$)} & \colhead{} & \colhead{}} 
\startdata
$34248.82$ & $3_{1,3}\rightarrow2_{1,2}$ & 7 & 4.3 & -6.4784 & 1.4 & 56 & 7.3 & clean transition  \\
$35915.48$ & $3_{1,2}\rightarrow2_{1,1}$ & 7 & 4.4 & -6.4164 & 1.4 & 58 & 7.6 & blended with HOCH$_2$CH$_2$OH \\
$45659.99$ & $4_{1,4}\rightarrow3_{1,3}$ & 7 & 6.0 & -6.0647 & 2.4 & 88 & 13.4 & clean transition  \\
$46737.73$ & $4_{0,4}\rightarrow3_{0,3}$ & 7 & 3.4 & -6.0064 & 2.6 & 122 & 8.6 & clean transition  \\
$71800.18$ & $6_{1,5}\rightarrow5_{1,4}$ & 13 & 11.3 & -5.4428 & 3.5 & 121 & 6.3 & clean transition \\ 
$81630.08$ & $7_{0,7}\rightarrow6_{0,6}$ & 15 & 11.8 & -5.2587 & 3.4 & 153 & 8.2 & slightly blended with CH$_3$CH$_2$CHO  \\                    
$83749.29$ & $7_{1,6}\rightarrow6_{1,5}$ & 15 & 14.8 & -5.2342 & 3.4 & 116 & 11.8 & slightly blended with CH$_3$COOH \\
$91251.84$ & $8_{1,8}\rightarrow7_{1,7}$ & 17 & 18.1 & -5.1166 & 1.7 & 104 & 11.2 & slightly blended with (CH$_2$OH)$_2$ and unidentified species  \\ 
$93505.09$ & $8_{2,7}\rightarrow7_{2,6}$ & 17 & 26.5 & -5.1061 & 1.4 & 46 & 6.0 & blended with CH$_3$\isotope[18]{O}H
\enddata
\tablenotetext{a}{$\mathrm{J^{''}_{K^{''}_{a}K^{''}_{c}}\rightarrow J^{'}_{K^{'}_{a}K^{'}_{c}}}$ (where the double prime indicates the upper state).}
\tablenotetext{c}{Signal to noise ratio is calculated from the integrated signal and noise level $\mathrm{\sigma=rms\,\sqrt{\delta v\,FWHM}}$, where $\mathrm{\delta v}$ is the velocity resolution of the spectra.}
\end{deluxetable*}

%In this way, the spectroscopic information of t-HC(O)SH was taken from its CDMS entry (62515), that is based on the previous studies from$\;$\cite{hocking1976rotationall} and$\;$\cite{hocking1976rotational}.
%This molecule is an asymmetric top in the nearly prolate limit $(\kappa=(2B-A-C)/(A-C)\approx-0.98)$.
%and thus, due to the difference between the inertia moments, the a-type transitions are expected to be detected in the ISM more easily than the b-type transitions.

The fitting procedure was performed considering the total emission of any other identified molecule in the spectral survey. This evaluation was carried out introducing all the compounds already detected in the ISM and in G+0.693 \citep[see][]{requenatorres2006,requena2008galactic,zeng2018complex,rivilla2019_cyanomehtanimine,ijimenez2020}. For the analysis, it was also assumed a cosmic microwave background temperature (T$_\mathrm{bg}$) of 2.73$\,$K and no background continuum source$\,$\citep{zeng2020}.

The global fit of all rotational lines to the observed data is shown in blue lines in Figure$\;$\ref{fig:monothioformic}, while in red we show the fit of the individual lines of t-HC(O)SH, and the observational data in black. As shown in both Figure$\;$\ref{fig:monothioformic} and Table$\;$\ref{tab:thioformicacid}, we have detected a total of nine a-type transitions in the 7$\,$mm and 3$\,$mm bands, each one of them detected above the 5$\sigma$ level in integrated intensity. Note that the SLIM synthetic spectrum shows small deviations with respect to the observations. These deviations are larger that the ones obtained for CH$_3$SH and g-C$_2$H$_5$SH (Appendix \ref{ap:A}). This is due to the fact that the t-HC(O)SH lines are significantly weaker than those of CH$_3$SH and g-C$_2$H$_5$SH (3-7$\,$mK versus 5-12$\,$mK and $>$40$\,$mK, respectively). However, the fit of all clean t-HC(O)SH lines is consistent with the noise. We also note that for the three slightly blended lines of t-HC(O)SH, the contribution of the other molecules is less than 10\% of the total intensity. Despite that the remaining two appear blended, the predicted LTE intensities are consistent with the observed lines. Note that the rest of the t-HC(O)SH lines covered within the observed frequency range are not shown due to strong blending issues.
%{\bf In Appendix X we added the remaining lines of t-HC(O)SH in the 3$\,$mm and 7$\,$mm bands.}

The physical parameters obtained from the fit are listed in Table$\;$\ref{tab:physical_parameters_all}. The values obtained for $T_\mathrm{ex}$ and $FWHM$ are consistent with those obtained previously for other molecular species toward this cloud \citep[$T_{\mathrm{ex}}\sim\,$5 - 20$\;$K and linewidths around 20 km s$^{-1}$;][]{zeng2018complex,rivilla2020prebiotic}. The low $T_{\mathrm{ex}}$ indicates that the emission of the molecules in G+0.693 is sub-thermally excited as a result of the low H$_2$ densities of this source \citep[$T_{\mathrm{kin}}\sim$150$\,$K;][]{requenatorres2006,zeng2018complex}.

For t-HC(O)SH, the fitted column density gives (1.6$\,\pm\,$0.1)$\times$10$^{13}$\,cm$^{-2}$. In addition, we derived a 3$\,\sigma$ upper limit of $\leq$3$\times$10$^{12}\,$cm$^{-2}$ for c-HC(O)SH (Table$\;$\ref{tab:physical_parameters_all}), as no clear transition has been detected within our dataset. For the calculation, we have assumed the same $T_{\mathrm{ex}}$, $v_{\mathrm{LSR}}$ and $FWHM$ as for t-HC(O)SH. A ratio c-HC(O)SH$\,$/$\,$t-HC(O)SH$\,\leq\,$0.2 was obtained.

%Table 2: physical parameters from LTE fit
\begin{deluxetable*}{cccccc}
    \tablecaption{Physical parameters of the species derived by LTE analysis in \textsc{madcuba}. \label{tab:physical_parameters_all}}
    \tablecolumns{6}
    \tablewidth{0pt}
    \tablehead{\colhead{Molecular formula} & \colhead{N} & \colhead{$\mathrm{T_{ex}}$} & \colhead{$\mathrm{v_{LSR}}$} & \colhead{FWHM} & \colhead{Abundance} \\
    \colhead{} & \colhead{$(\times$ 10$^{13}$ cm$^{-2})$} & \colhead{(K)} & \colhead{$\mathrm{(km\,s^{-1})}$} & \colhead{$\mathrm{(km\,s^{-1})}$} & \colhead{($\tfrac{N(X)}{N(H_2)}\times10^{-10}$)}}
\startdata
    $\mathbf{trans-HC(O)SH}$ & $1.6 \pm 0.1$ & $10 \pm 1$ & $69.0^{\,\mathrm{a}}$ & $21.0^{\,\mathrm{a}}$ & $1.2\pm 0.2$\\
    $\mathbf{cis-HC(O)SH}$ & $\leq 0.3$ & $10.0$ & $69.0^{\,\mathrm{a}}$ & $21.0$\tablenotemark{a} & $\leq 0.2$\\
    $\mathbf{g-C_2H_5SH}$ & $4\pm2$ & $10\pm 5$ & $69.0^{\,\mathrm{a}}$ & $20.0$\tablenotemark{a} & $3\pm1$ \\
   $\mathbf{a-C_2H_5SH}$ & $\leq 2.3$ & $9.91 $ & $69.0^{\,\mathrm{a}}$ & $20.0^{\,\mathrm{a}}$ & $\leq 1.7$\\
    $\mathrm{CH_3SH,\, K_a=0}$ & $46.8\pm0.5$ & $8.5\pm0.1$ & $68.0\pm0.1$ & $21.2\pm0.3$ & -\\
    $\mathrm{CH_3SH,\, K_a=1}$ & $18.6\pm 0.7$ & $14.9\pm0.5$ & $68.8\pm0.3$ & $22.0\pm0.7$ & -\\
    $\mathbf{CH_3SH}$ & $65\pm2$ & - & - & - & $48\pm 5$\\
%   
   % $\mathbf{t-HC(S)SH}$ & $\leq 0.6$ & 10.7 & 69.0 & 21.0 & $\leq 0.4$\\
%
    $\mathbf{t-HCOOH}$ & $20\pm4$ & $10\pm2$ & $68\pm2$ & $22\pm5$ & $15\pm4$ \\
\enddata
    \tablenotetext{a}{Value fixed in the fit.}
\end{deluxetable*}

\subsection{Detection of g-C$_2$H$_5$SH and CH$_3$SH}

%C$_2$H$_5$SH shows two relevant conformers: synclinal and antiperiplanar, associated with the rotation of the S-C bond. The former one is double degenerate (conformer + and -) and is $\sim$ 160 K lower in energy than the anti conformer$\,$\citep{durig1975raman}. 

This molecule was tentatively detected in Orion KL \citep{kolesnikova2014}, based on a few isolated transitions assigned to this isomer. We report here an unambiguous detection of this isomer that confirms it presence in the ISM. We have listed in Table$\;$\ref{tab:gC2H5SH} (Appendix A) the transitions measured toward G+0.693. Note that eight of the targeted lines are totally clean (Figure$\;$\ref{fig:c2h5sh}, Appendix A) and above the 5$\sigma$ level in integrated intensity. We derived a total column density of (4$\,\pm\,$2)$\times$10$^{13}\,$cm$^{-2}$ (Table$\;$\ref{tab:physical_parameters_all}).

%The spectroscopic information of the rest frequencies for the transitions of CH$_3$SH was taken from the CDMS entry (048510). 
In the case of CH$_3$SH, the the fit was carried out separating the targeted lines into its $K_a=0$ and $K_a=1$ levels\footnote{The moments of inertia of CH$_3$SH frame it in the limiting prolate case, $\kappa=(2B-A-C)/(A-C)\approx-0.988$.}.
%Hence, the fit was carried out separating the targeted lines into its $K_a=0$ and $K_a=1$ levels. 
%For each group of transitions, the \textsc{slim-autofit} tool was employed.
The brightest transitions fall into the 3$\,$mm and 2$\,$mm bands as presented in Figure$\;$\ref{fig:ch3sh} and Table$\;$\ref{tab:ch3sh} (Appendix B). Note that the agreement between the predicted and observed spectra is excellent for all clean transitions. The column density using the $K_a=0$ and $K_a=1$ levels is (6.5$\,\pm\,$0.2)$\times$10$^{14}\,$cm$^{-2}$ (Table$\;$\ref{tab:physical_parameters_all}). This gives a ratio of CH$_3$SH$\,$/$\,$g-C$_2$H$_5$SH$\;$=$\;$16$\,\pm\,$7. Note that the column density of $K_a=2$ and remaining transitions in CH$_3$SH contribute less than a 10\% in the total determined. 

Finally, we note that both \isotope[13]{C} and \isotope[34]{S} isotopologues of CH$_3$SH have also been detected with a few clean lines and will be presented in a forthcoming paper (Colzi et al. in prep.).

\subsection{Molecular Abundances and comparison with their O-bearing analogues} \label{sec:abundances}

To derive the fractional abundances of these species relative to H$_2$, we have assumed an H$_2$ column density of $N_{H_2}=1.35\times10^{23}\,$cm$^{-2}$ \citep[][]{martin2008}. This gives  $\sim\,$3$\,\times\,$10$^{-10}$, $\sim\,$1$\,\times\,$10$^{-10}$ and $\sim\,$5$\,\times\,$10$^{-9}$ for g-C$_2$H$_5$SH, t-HC(O)SH and CH$_3$SH, respectively (Table$\,$\ref{tab:physical_parameters_all}). 

\begin{figure}[htp!]
    \centering
    \includegraphics[width=0.45\textwidth]{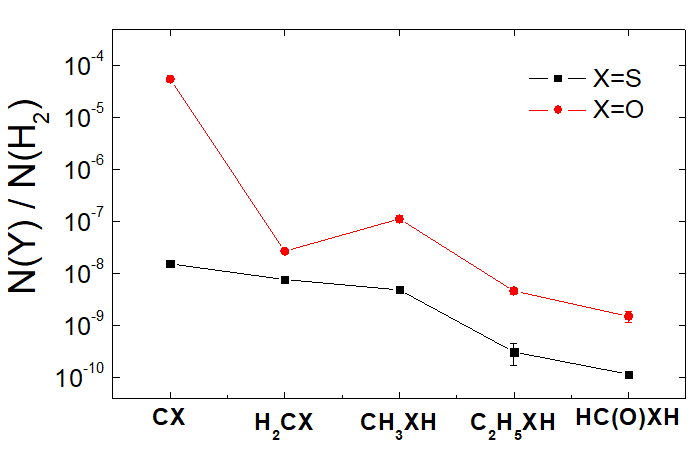}
    \caption{Molecular abundances wrt H$_2$ for the reported detections and their OH-analogues (CH$_3$OH, C$_2$H$_5$OH and HC(O)OH) including CS, CO, H$_2$CO and H$_2$CS. The lines that connect each dot do not have any physical meaning but are just a visual aid.}
    \label{fig:fractional_abundances}
\end{figure}

In Figure$\;$\ref{fig:fractional_abundances} we have plotted these values and compared them with the abundances obtained for their OH molecular analogues, namely: C$_2$H$_5$OH (ethanol), HC(O)OH (formic acid) and CH$_3$OH (methanol). We obtained a ratio CH$_3$OH$\,$/$\,$CH$_3$SH$\,$=$\,$23, C$_2$H$_5$OH$\,$/$\,$C$_2$H$_5$SH$\,$=$\,$15 and HC(O)OH$\,$/$\,$HC(O)SH$\,$=$\,$13. Although these OH-analogues are more abundant by a factor$\,\geq$10, the resulting trend is strikingly similar. Note, however, that this trend is lost when we compare molecules such as carbon monosulfide (CS) and thioformaldehyde (H$_2$CS) with their O-bearing analogues, resulting in H$_2$CO$\,$/$\,$H$_2$CS$\,$=$\,$3.5 and CO$\,$/$\,$CS$\,$=$\,$3.5$\times$10$^3$ (Figure \ref{fig:fractional_abundances}).

%Figure 2: comparison OH-analogues 

Since CH$_3$OH, C$_2$H$_5$OH, H$_2$CO, CO and CS are optically thick toward G+0.693, the column density of these molecules were inferred from  CH$_3$\isotope[18]{O}H, \isotope[13]{C}H$_3$CH$_2$OH, H$_2$C\isotope[18]{O}, C\isotope[18]{O} and \isotope[13]{C}\isotope[34]{S}. It was assumed \isotope[16]{O}$\,$/$\,$\isotope[18]{O}$\,$=$\,$250$\;$ and  \isotope[12]{C}$\,$/$\,$\isotope[13]{C}$\,$=$\,$21 \citep[][]{armijos2014} and \isotope[32]{S}$\,$/$\,$\isotope[34]{S}$\,$=$\,$22 \citep[][]{wilson1999}. For HC(O)OH, the main isotopologue of HC(O)OH was employed as the ratio H\isotope[12]{C}(O)OH$\,$/ H\isotope[13]{C}(O)OH is consistent with the \isotope[12]{C}$\,$/$\,$\isotope[13]{C} ratio.
%measured by \citet{armijos2014}. 
For the sulfur-analogues, except CS, we do not expect opacity issues due to their much lower abundances. The information about the fit employed for the molecules presented in Figure \ref{fig:fractional_abundances} or their isotopologues is included in Appendix \ref{ap:B}.

%\subsection{Isomeric ratios}

%Only the lower-energy rotamers of HC(O)SH and C$_2$H$_5$SH have been detected toward G+0.693. This is in agreement with the so-called minimum energy principle \citep[MEP;][]{lattelais2009}, which states that the thermodynamically favored isomers are more abundant in the ISM. %This is consistent with the detections we are reporting in this work, since in both cases we have only found the lower-energy rotamer: t-HC(O)SH and g-C$_2$H$_5$SH. 

%According to our analysis, a ratio $\leq 0.2$ and $\leq 0.5$ was obtained for c-HC(O)SH$\,$/$\,$t-HC(O)SH and a-C$_2$H$_5$SH$\,$/$\,$g-C$_2$H$_5$SH. In each case, the value is still consistent with the thermodynamic equilibrium value at $T_{\mathrm{K}}\,$=$\,$100$\,$K ($\sim\,$0.04 and $\sim\,$0.1, respectively). Even though, for a deeper analysis, the detections of such isomers is required \citep[as it was previously done, for example, with the Z and E isomers of cyanomethanimine;][]{rivilla2019_cyanomehtanimine}.

%Although we do not have information about the barriers involved in these processes, as in G+0.693 the kinetic temperatures ($T_{\mathrm{K}}$) range from 50$\,$K to 140$\,$K \citep[][]{zeng2018complex} the isomerization in both cases would probably be hampered. 

\section{Discussion}
\label{sec:discussion}

\subsection{Comparison with previous observations}

We have compared different column density ratios obtained between the S-bearing compounds measured toward G+0.693 and two other sources, namely: Orion KL \citep[][]{kolesnikova2014} and Sgr B2(N2) \citep[][see Table \ref{tab:abundances}]{muller2016}. 
%In contrast with G+0.693, these sources are star-forming regions which show higher temperatures ($\geq\,$100$\,$K) and densities \citep[$\geq\,$10$^6\,$cm$^{-3}$;][]{feng2015,bonfand2019}.

From this table, we find that all abundance ratios measured in G+0.693 are strikingly similar. For instance, when we compare SH-bearing molecules with their OH-analogues (CH$_3$OH / CH$_3$SH and C$_2$H$_5$OH / C$_2$H$_5$SH) we recover the trend already found in Figure \ref{fig:fractional_abundances}, which might indicate a similar chemistry between OH- and SH-bearing species. 
The CH$_3$OH$\,$/$\,$CH$_3$SH ratio measured in G+0.693 is a factor of 5 lower than those found in Sgr B2(N2) and Orion KL, which indicates that G+0.693 is richer in sulfur-bearing species than the two massive hot cores. This may be related to the fact that the chemistry of this cloud is affected by large-scale shocks \citep[][]{requenatorres2006,zeng2018complex}. Since sulfur is heavily depleted on grains \citep[possibly in the form of S$_8$ and other sulfur allotropes;][]{shingledecker2020sulphur}, the sputtering of dust grains induced by shocks could liberate a significant fraction of the locked sulfur.

%This could be due to a different sulfur chemistry in G+0.693 that leads to an extremely efficient production of CH$_3$SH.
The C$_2$H$_5$OH$\,$/$\,$C$_2$H$_5$SH ratio is comparable to the value determined toward Orion KL while there is a difference by a factor of 10 between G+0.693 and the upper limit toward Sgr B2(N2). All these ratios suggest that sulfur is less incorporated into molecules than oxygen, which again might be due to the fact that a significant fraction of sulfur is locked up in grains.

%Table 3: abundances
\begin{deluxetable*}{cccccc}[htp!]
    \tablecaption{Relative abundances of molecules and comparison with other sources.\label{tab:abundances}}
    \tablecolumns{5}
    \tablewidth{0pt}
    \tablehead{
    \colhead{Source} &
    \colhead{CH$_3$SH$\,$/$\,$C$_2$H$_5$SH} &
    \colhead{CH$_3$OH$\,$/$\,$C$_2$H$_5$OH} &
    \colhead{CH$_3$OH$\,$/$\,$CH$_3$SH} &
    \colhead{C$_2$H$_5$OH$\,$/$\,$C$_2$H$_5$SH} &
    }
    \startdata
    G+0.693 & 16$\pm$7 & 24$\pm$4\tablenotemark{c,d} & 23$\pm$3\tablenotemark{c} & 15$\pm$7\tablenotemark{d} \\
    Sgr B2(N2)\tablenotemark{a} & $\geq$21 & 20 & 118 & $\geq$125 \\
    Orion KL\tablenotemark{b} & 5 & 31 & 120 & 20 \\
    \enddata
    \tablenotetext{a}{Data taken from \citet[][]{muller2016}.}
    \tablenotetext{b}{Data taken from \citet[][]{kolesnikova2014}.}
    \tablenotetext{c}{Data inferred from CH$_3$\isotope[18]{O}H assuming \isotope[16]{O} / \isotope[18]{O} = 250 \citep[][]{armijos2014}.}
    \tablenotetext{d}{Data inferred from \isotope[13]{C}H$_3$CH$_2$OH assuming \isotope[12]{C} / \isotope[13]{C} = 21 \citep{armijos2014}.}
\end{deluxetable*}

%As for any case, it is necessary to point out the possible errors made in the calculation of these ratios, despite we have made sure the validity of the assumptions we made.  
% de qué manera está justificado el uso del isotopólogo 18O en CH3OH y 13C en C2H5OH?
%  CH$_3$OH / CH$_3$SH and C$_2$H$_5$OH / C$_2$H$_5$SH

\subsection{Interstellar chemistry of thiols}
\label{sec:chemicalpathways}

In order to understand the chemistry of S-bearing compounds, different models have been proposed based on the simple molecules detected in the gas phase \citep[][]{muller2016,gorai2017,lamberts2018,laas2019}. 
%{\bf In addition, only two S-bearing species (OCS and SO$_2$) have been detected in interstellar ices \citep[][]{palumbo1997}.}
CH$_3$SH is thought to be formed on grain surfaces by sequential hydrogenations starting from CS:
$$
\mathrm{CS + H \rightarrow HCS }
$$
$$
\mathrm{HCS + H \rightarrow H_2CS}
$$
$$
\mathrm{H_2CS + H \rightarrow CH_3S}
$$
$$
\mathrm{CH_3S + H \rightarrow CH_3SH.}
$$

The last step could also yield other products but theoretical calculations show a branching ratio (br) of $\sim$75\% for CH$_3$SH \citep[][]{lamberts2018}. This is a viable mechanism since approximately half of the CS present in the ices is available to undergo hydrogenation, while the other half is converted to OCS \citep[][]{palumbo1997}. Once formed, CH$_3$SH could remain stored in the ices until released by grain sputtering in the large-scale shocks present in G+0.693 \citep[][]{requenatorres2006,zeng2020}.
%Regardless, in contrast with CO, the hydrogenation route of CS does not seem to explain the abundances observed in Figure$\;$\ref{fig:fractional_abundances}, as CH$_3$SH appear to be slightly more abundant when compared to its precursors.

Likewise, ethyl mercaptan is proposed to be formed by radical-radical reactions \citep[][]{gorai2017,muller2016} such as:
$$
\mathrm{C_2H_5 + SH \rightarrow C_2H_5SH}
$$
$$
\mathrm{CH_3 + CH_2SH \rightarrow C_2H_5SH.}
$$

Observations of the reactants could give us a hint of the dominant reaction based on the measured column densities. However, chemical modeling is needed to understand the efficiency of these formation routes. Note that CH$_2$SH is the main product of the hydrogenation of H$_2$CS \citep[][]{lamberts2018}.

To our knowledge, there is no information available in the literature about the chemistry of HC(O)SH. However, we can make a guess and assume similar astrochemical pathways for the SH-based species as the ones for their OH-analogues. A possible pathway to HC(O)SH could mimic the formation of HC(O)OH. \citet[][]{ioppolo2011} showed that the formation of HC(O)OH starts from CO and the OH radical in the ice. The thiol-equivalent reaction would be:
$$
\mathrm{CO + SH \rightarrow HSCO \xrightarrow[]{\text{H}} HC(O)SH.}
$$

The first step has been found to be efficient by \citet[][]{adriens2010}, but further experimental and/or theoretical work is needed to investigate whether HSCO could be hydrogenated further.

Another possibility could be:
$$
\mathrm{HCO + SH \rightarrow HCOSH}
$$
or 
$$
\mathrm{OCS \xrightarrow[]{2H}  HC(O)SH.}
$$

The first reaction was initially proposed for HC(O)OH by \citet[][]{garrod2006} and could be a viable mechanism since SH can be formed on surfaces from S + H or H + H$_2$S via tunneling \citep[][]{vidal2017}. The second involves the sequential hydrogenation of OCS, which is a molecule detected in ices \citep[][]{palumbo1997} and it is moderately abundant in the gas phase in G+0.693 \citep[N$>$10$^{15}$cm$^{-2}$;][]{armijos2014}. Theoretical studies about these chemical networks are currently under work and will be presented in a forthcoming paper (Molpeceres et al. in prep).
%Although the kinetic and thermodynamical properties of this reaction are unknown, the high abundance of OCS in G+0.693 \citep[log N $>$ 15;][]{armijos2014} makes it one plausible chemical precursor.  

\subsection{Implications for theories on the origin of life}

As it was previously stated, thioacids and thioesters have been proposed as key agents in the polymerization of amino acids into peptides and proteins \citep{foden2020prebiotic,muchowska2020}. In some of these works, it is stressed the importance of cystein, HSCH$_2$CH(NH$_2$)COOH, as the primary organic source of sulfide in biology and a key catalyst in the abiotic polymerization of peptides. Smaller thiols such as CH$_3$SH and C$_2$H$_5$SH are also believed to play a key role in prebiotic chemistry and in theories about the origin of life, since they are precursors for the synthesis of the aminoacids methionine and ethionine \citep[][]{parker2011}. In turn, HC(O)SH could be an important ingredient in the phosphorylation of nucleosides as demonstrated by \citet{lohrmann1968}.

%{\bf Here continue with the discussion maybe mentioning the presence of SH-bearing molecules in meteorites and carbonaceous chondrites?? Make the link with the detection of CH3SH in comet 67P. What is the origin of these S-bearing complex organics? The ISM and then you continue with the following two paragraphs. 
The idea of an extraterrestrial delivery of SH-compounds onto Earth, is supported by the detection of CH$_3$SH in Murchinson carbonaceous chondrite \citep[][]{tingle1991} and in the coma of the 67P/Churyumov-Gerasimenko comet \citep[][]{calmonte2016}. In the latter, C$_2$H$_6$S was also detected (either in the form of C$_2$H$_5$SH or CH$_3$SCH$_3$ -dimethyl sulfide-) with an abundance ratio CH$_3$SH$\,$/$\,$C$_2$H$_6$S$\,\sim\,$10 which is consistent with the value obtained in G+0.693 (Table$\;$\ref{tab:abundances}). This resemblance could be possibly indicating a pre-solar origin of these compounds. Still, further studies are required within other regions in the ISM and planetary bodies to make a proper connection.

In summary, not only have our observations confirmed the presence of sulfur-bearing complex organics such as CH$_3$SH and C$_2$H$_5$SH, but they also have revealed the existence of the simplest thioacid known, HC(O)SH, in the ISM. 
%This represents a higher level in chemical complexity for S-bearing species since it involves an aldehyde group (-C(O)H).

%\section{Conclusions} \label{sec:conclusions}

%In this Letter we have reported the detection first thiol acid toward G+0.693. Additionally, clear detections of CH$_3$SH and C$_2$H$_5$SH have been presented with a review of their possible chemical routes of synthesis. We have also obtained a similar trend for the SH-bearing compounds studied and their OH-analogues. This could be revealing a similar chemical behavior between these compounds and/or an unknown chemistry happening toward G+0.693 that leads to an extremely efficient production of CH$_3$SH. Alternatively, a link with these compounds in the context of prebiotic chemistry has been reviewed.

\acknowledgments
\section{Acknowledgments}

L.F.R.-A. acknowledges support from a JAE-intro ICU studentship funded by the Spanish National Research Council (CSIC). L.F.R.-A., V.M.R. and L.C. also acknowledge support from the Comunidad de Madrid through the Atracci\'on de Talento Investigador Modalidad 1 (Doctores con experiencia) Grant (COOL: Cosmic Origins Of Life; 2019-T1/TIC-15379; PI: V.M.Rivilla). I.J.-S. and J.M.-P. have received partial support from the State Research Agency (AEI) through project numbers PID2019-105552RB-C41 and MDM-2017-0737 Unidad de Excelencia "María de Maeztu" - Centro de Astrobiología (CSIC-INTA). PdV and BT thank the support from the European Research Council through Synergy Grant ERC-2013-SyG, G.A. 610256 (NANOCOSMOS) and from the Spanish Ministerio de Ciencia e Innovación (MICIU) through project PID2019-107115GB-C21. BT also thanks the Spanish MICIU for funding support from grants AYA2016-75066-C2-1-P and PID2019-106235GB-I00.


\vspace{5mm}
\facilities{IRAM 30m, Yebes 40m}


\software{\textsc{madcuba}}



%-----APPENDICES

\appendix
%-----APPENDIX A. thioformic and ethyl tables

\section{Detections of C$_2$H$_5$SH and CH$_3$SH in G+0.693}
\label{ap:A}

\begin{figure*}[h]
    \centering
    \includegraphics[width=1\textwidth]{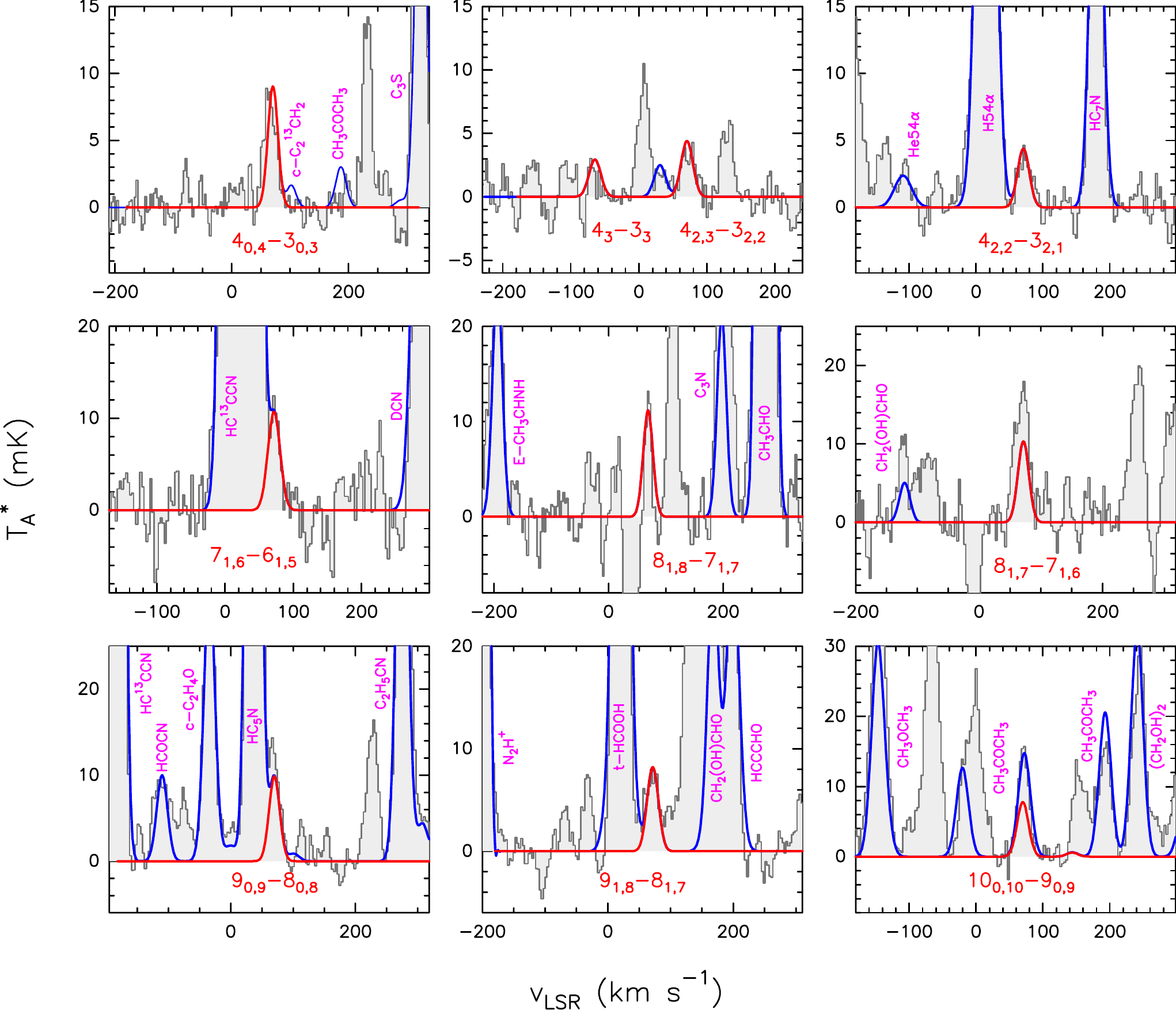}
    \caption{Cleanest and brightest lines of g-C$_2$H$_5$SH detected toward G+0.693. The red line shows the best LTE fit to the observed spectra (represented by the black lines) with their corresponding quantum numbers in red ( J$_{\mathrm{K_a,K_c}}$; see Table \ref{tab:gC2H5SH} for a description of their tunneling states). The data has been smoothed up to 3 km s$^{-1}$ for an optimal line visualization. The blue lines show the spectra including the emission of all the molecules searched toward the cloud. Note that these lines are tagged with their corresponding molecular compound in pink.}
    \label{fig:c2h5sh}
\end{figure*}

\begin{figure*}[h]
    \centering
    \includegraphics[width=1\textwidth]{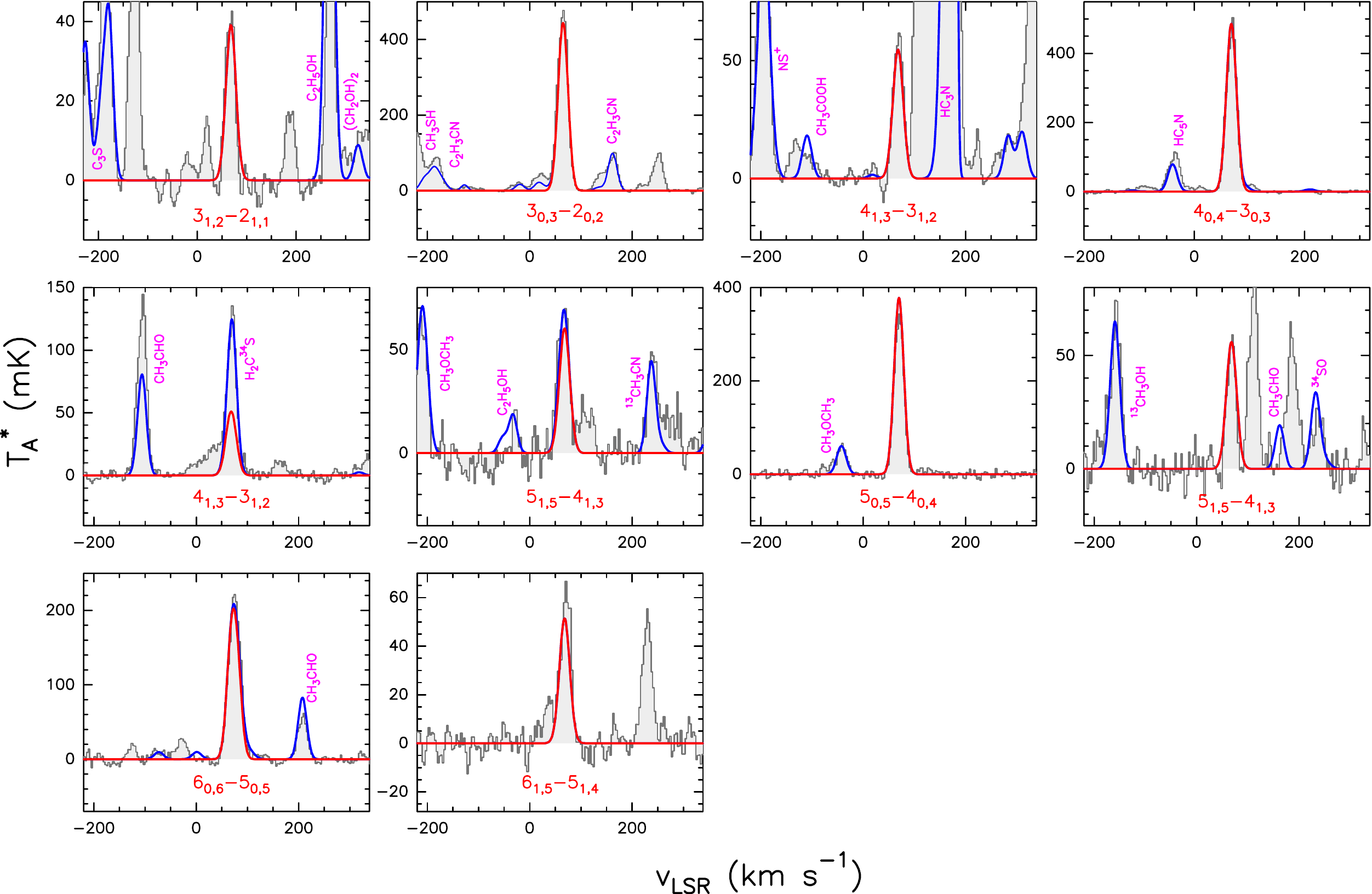}
    \caption{Cleanest and brightest lines of CH$_3$SH detected toward G+0.693. The red line shows the best LTE fit to the observed spectra (represented by the black lines) with their corresponding quantum numbers in red ( J$_{\mathrm{K_a,K_c}}$; see Table \ref{tab:ch3sh} for the description of their torsional states). The data has been smoothed up to 3 km s$^{-1}$ for an optimal line visualization. The blue lines show the spectra including the emission of all the molecules searched toward the cloud. Note that these lines are tagged with their corresponding molecular compound in pink.}
    \label{fig:ch3sh}
\end{figure*}

In Figures \ref{fig:c2h5sh} and \ref{fig:ch3sh} and Tables \ref{tab:gC2H5SH} and \ref{tab:ch3sh} the fit of g-C$_2$H$_5$SH and CH$_3$SH is shown. The fitting procedure followed has been the same as t-HC(O)SH, explained in the main text.

%Figure 2: transitions of C2H5SH

%C2H5SH
\begin{deluxetable*}{cccccccccl}[h]
    \tablecaption{Lines of g-C$_2$H$_5$SH detected toward G+0.693 with their corresponding QNs and tunneling states (TS), $\mathrm{\log A_{ul}}$, $\mathrm{g_u}$ and $\mathrm{E_u}$. $\int T_A^* d\nu$ and rms are also provided and used to calculate the SNR of the detected lines.\label{tab:gC2H5SH}}
    \tablecolumns{10}
    \tablewidth{0pt}
    \tablehead{
    \colhead{Rest frequency} & \colhead{QNs} & TS & \colhead{g$_u$} & \colhead{E$_u$} & \colhead{$\mathrm{\log A_{ul}}$} &     \colhead{rms}& \colhead{$\int T_A^* d\nu$} & \colhead{SNR} & \colhead{Comments} \\
    \colhead{(MHz)} & \colhead{} & \colhead{$\nu''\rightarrow\nu'$} & \colhead{} & \colhead{(K)} & \colhead{(s$^{-1}$)} & \colhead{(mK)} &   \colhead{($\mathrm{mK\,km\,s^{-1}}$)} & \colhead{} & \colhead{}}  
    \startdata
% $31092.310$ & $3_{1,2}\rightarrow2_{1,1}$ & 7 & 2.71 & -6.5274 & 1.7 & 61 & 6.6 & clean transition  \\ 
 % $31101.080$ & $3_{1,2}\rightarrow2_{1,1}$ & 7 & 2.63 & -6.5392 & 1.7 & 61 & 6.6 & blended with CH$_3$COCH$_3$ \\ 
 %   \addlinespace
    $40499.173$ & $4_{0,4}\rightarrow3_{0,3}$ & $1\rightarrow1$ & 9 & 3.0 & -6.1161 & 1.4 & 225 & 29 & clean transition\tablenotemark{a}  \\ 
 %   \addlinespace
    $40499.591$ & $4_{0,4}\rightarrow3_{0,3}$ & $0\rightarrow0$ & 9 & 2.9 & -6.1163 & 1.4 & - & - & -  \\ 
 %   \addlinespace
    $40558.849$ & $4_{2,3}\rightarrow3_{2,2}$ & $1\rightarrow1$ & 9 & 7.6 & -6.2391 & 1.2 & 107 & 16 & clean transition\tablenotemark{a}  \\ 
 %   \addlinespace
    $40559.284$ & $4_{2,3}\rightarrow3_{2,2}$ & $0\rightarrow0$ & 9 & 7.5 & -6.2392 & 1.2 &  - & - & - \\ 
 %   \addlinespace
    $40576.963$ & $4_{3,2}\rightarrow3_{3,1}$ & $1\rightarrow1$ & 9 & 13.2 & -6.2004 & 1.2 & 72 & 11 & clean transition\tablenotemark{a}  \\ 
 %   \addlinespace
    $40577.299$ & $4_{3,1}\rightarrow3_{3,0}$ & $1\rightarrow1$ & 9 & 13.2 & -6.2004 & 1.2 & - & - & -  \\ 
 %   \addlinespace
    $40577.432$ & $4_{3,2}\rightarrow3_{3,1}$ & $0\rightarrow0$ & 9 & 13.2 & -6.2004 & 1.2 & - & - & -  \\ 
 %   \addlinespace
    $40577.769$ & $4_{3,1}\rightarrow3_{3,0}$ & $0\rightarrow0$ & 9 & 13.2 & -6.2004 & 1.2 & - & - & -  \\ 
 %   \addlinespace
    $40622.316$ & $4_{2,2}\rightarrow3_{2,1}$ & $1\rightarrow1$ & 9 & 7.6 & -6.2370 & 1.2 & 107 & 16 & clean transition\tablenotemark{a}  \\ 
 %   \addlinespace
    $40622.792$ & $4_{2,2}\rightarrow3_{2,1}$ & $0\rightarrow0$ & 9 & 7.5 & -6.2371 & 1.2 & - & - & -  \\ 
 %   \addlinespace
    $72466.861$ & $7_{1,6}\rightarrow6_{1,5}$ & $1\rightarrow1$ & 15 & 11.7 & -5.3458 & 2.8 & 261 & 17 & clean transition\tablenotemark{a}  \\
    $72468.152$ & $7_{1,6}\rightarrow6_{1,5}$ & $0\rightarrow0$ & 15 & 11.6 & -5.3459 & 2.8 & - & - & -  \\ 
 %   \addlinespace
 % 
    $79204.501$ & $8_{1,8}\rightarrow7_{1,7}$ & $1\rightarrow1$ & 17 & 14.6 & -5.2243 & 1.8 & 204 & 24 & clean transition\tablenotemark{a}  \\
   $79204.501$ & $8_{1,8}\rightarrow7_{1,7}$ & $0\rightarrow0$ & 17 & 14.6 & -5.2243 & 1.8 & - & - & -  \\
  $82782.441$ & $8_{1,7}\rightarrow7_{1,6}$ & $1\rightarrow1$ & 17 & 15.2 & -5.1667 & 1.8 & 233 & 30 & slightly blended with unidentified line\tablenotemark{a}  \\ 
 % esta también
  $82783.912$ & $8_{1,7}\rightarrow7_{1,6}$ & $0\rightarrow0$ & 17 & 15.1 & -5.1668 & 1.4 & - & - & -  \\  
 %   \addlinespace
  %  $89067.529$ & $9_{1,9}\rightarrow8_{1,8}$ & 19 & 18.27 & -5.0672 & 1.2 & 101 & 15.4 & blended with C$_3$N\tablenotemark{a} \\ 
 %   \addlinespace
 %   $89067.529$ & $9_{1,9}\rightarrow8_{1,8}$ & 19 & 18.35 & -5.067 & 1.2 & 100 & 15.2 & - \\ 
 %   \addlinespace
    $90516.341$ & $9_{0,9}\rightarrow8_{0,8}$ & $1\rightarrow1$ & 19 & 17.6 & -5.0412 & 1.2 & 223 & 34 & slightly blended with unidentified line\tablenotemark{a}   \\ 
    $90516.926$ & $9_{0,9}\rightarrow8_{0,8}$ & $0\rightarrow0$ & 19 & 17.5 & -5.0414 & 1.2 & - & - & -   \\ 
  %  \addlinespace
    $93082.083$ & $9_{1,8}\rightarrow8_{1,7}$ & $1\rightarrow1$ & 19 & 19.1 & -5.0096 & 2.0 & 191 & 18 & clean transition\tablenotemark{a}  \\
   $93083.720$ & $9_{1,8}\rightarrow8_{1,7}$ & $0\rightarrow0$ & 19 & 19.1 & -5.0097 & 2.0 & - & - & -  \\ 
  %  \addlinespace
   %   \addlinespace
    $100391.240$ & $10_{0,10}\rightarrow9_{0,9}$ & $1\rightarrow1$ & 21 & 21.9 & -4.9042 & 1.8 & 171 & 17 & blended with CH$_3$COCH$_3$\tablenotemark{a}  \\ 
  %  \addlinespace
    $100391.791$ & $10_{0,10}\rightarrow9_{0,9}$ & $0\rightarrow0$ & 21 & 21.8 & -4.9042 & 1.8 & - & - & - \\ 
 %  $103364.095$ & $10_{1,9}\rightarrow9_{1,8}$ & 21 & 23.60 & -4.8698 & 2.7 & 72 & 4.9 & slightly blended with unidentified transition  \\ 
 %  $103364.880$ & $10_{1,9}\rightarrow9_{1,8}$ & 21 & 23.52 & -4.8700 & 2.7 & 73 & 4.9 & slightly blended with unidentified transition   \\ 
    \enddata
    \tablenotetext{a}{In these cases, the lines are composed by a blend of individual g-C$_2$H$_5$SH lines. Here, SNR is calculated according with Figure \ref{fig:c2h5sh}.}
\end{deluxetable*}

\newpage

%Table 4: transitions of CH3SH 
\begin{deluxetable*}{cccccccccl}[h]
    \tablecaption{Lines of CH$_3$SH detected toward G+0.693 with their corresponding quantum numbers (QNs) and torsional symmetry (T$_\mathrm{s}$), logarithm of the Einstein coefficients ($\mathrm{\log A_{ul}}$), degeneracy ($\mathrm{g_u}$) and energy ($\mathrm{E_u}$) of the upper state. The peak intensity ($\mathrm{I_{peak}}$) and root mean square (rms) noise level are also provided and used to calculate the signal to noise ratio (SNR) of the detected lines.\label{tab:ch3sh}}
    \tablecolumns{10}
    \tablewidth{0pt}
    \tablehead{
    \colhead{Rest frequency} & \colhead{QNs} & T$_\mathrm{s}$ & \colhead{g$_u$} & \colhead{E$_u$} & \colhead{$\mathrm{\log A_{ul}}$} & \colhead{rms}& \colhead{$\mathrm{I_{peak}}$} & \colhead{SNR} & \colhead{Comments} \\
    \colhead{(MHz)} & \colhead{} & \colhead{} & \colhead{} &\colhead{(K)} & \colhead{(s$^{-1}$)} & \colhead{(mK)} & \colhead{(mK)} & \colhead{} & \colhead{}}  
    \startdata
    75085.898 & $3_{1,2}\rightarrow2_{1,1}$ & A$^+$ & 7 & 8.7 & -5.5086 & 2.7 & 39 & 14 & clean transition \\
    75862.889 & $3_{0,3}\rightarrow2_{0,2}$ & A$^+$ & 7 & 3.6 & -5.4444 & 2.3 & 260 & 113 & clean transition \\
    75864.422 & $3_{0,3}\rightarrow2_{0,2}$ & E & 7 & 5.1 & -5.4443 & 2.3 & 221 & 96 & clean transition  \\
    100110.219 & $4_{1,3}\rightarrow3_{1,2}$ & A$^+$ & 9 & 12.3 & -5.0966 & 3.1 & 55 & 18 & clean transition \\
    101139.150 & $4_{0,4}\rightarrow3_{0,3}$ & A$^+$ & 9 & 7.3 & -5.0554  & 2.9 & 271 & 93 & clean transition \\
    101139.655 & $4_{0,4}\rightarrow3_{0,3}$ & E & 9 & 8.7 & -5.0543 & 2.9 & 230 & 79 & clean transition \\
    101284.366 & $4_{1,3}\rightarrow3_{1,2}$ & E & 9 & 13.5 & -5.0818 & 2.9 & 51 & 18 & blended with $\mathrm{H_2C^{34}S}$ \\
    125130.863 & $5_{1,5}\rightarrow4_{1,3}$ & A$^+$ & 11 & 17.1 & -4.7869 & 6.2 & 61 & 10 & slightly blended with $\mathrm{HC_5N}$ \\
    126403.834 & $5_{0,5}\rightarrow4_{0,4}$ & E & 11 & 13.6 & -4.7556 & 6.4 & 183 & 29 & clean transition  \\
    126405.676 & $5_{0,5}\rightarrow4_{0,4}$ & A$^+$ & 11 & 12.2 & -4.7556  & 6.4 & 215 & 34 & clean transition  \\
    126683.419 & $5_{1,5}\rightarrow4_{1,9}$ & A$^+$ & 11 & 18.4 & -4.7707 & 6.5 & 56 & 9 & clean transition \\
    151654.218 & $6_{0,6}\rightarrow5_{0,5}$ & E & 13 & 19.6 & -4.5123 & 5.4 & 116 & 22 & clean transition \\
    151660.047 & $6_{0,6}\rightarrow5_{0,5}$ & A$^+$ & 13 & 18.2 & -4.5123 & 5.4 & 136 & 25 & clean transition  \\
    152129.018 & $6_{1,5}\rightarrow5_{1,4}$ & E & 13 & 24.4 & -4.5210 & 4.5 & 52 & 12 & clean transition \\
    \enddata
\end{deluxetable*}

\clearpage
%------APPENDIX B. isotopologues

\section{Figures and tables of \isotope[13]{C}\isotope[34]{S}, H$_2$CS, H$_2$C\isotope[18]{O}, CH$_3$\isotope[18]{O}H, \isotope[13]{C}H$_3$CH$_2$OH}
\label{ap:B}

We show a few intense the lines selected within the targeted frequency coverage in G+0.693 for some of the species used in the molecular abundances calculations in Figure \ref{fig:fractional_abundances}. Note that other visible lines not shown in these figures or tables are either heavily blended or in a bad part of the spectra.
%table 13C34S
\begin{deluxetable*}{cccccccccl}[h]
    \tablecaption{Lines of \isotope[13]{C}\isotope[34]{S} detected toward G+0.693 with their corresponding spectroscopic information and parameters derived from the LTE fit.\label{tab:13c34s}}
    \tablecolumns{9}
    \tablewidth{0pt}
    \tablehead{
    \colhead{Rest frequency} & \colhead{QNs} &  \colhead{g$_u$} & \colhead{E$_u$} & \colhead{$\mathrm{\log A_{ul}}$} & \colhead{rms}& \colhead{$\int T_A^* d\nu$} & \colhead{SNR} & \colhead{Comments} \\
    \colhead{(MHz)} & \colhead{} & \colhead{} &\colhead{(K)} & \colhead{(s$^{-1}$)} & \colhead{(mK)} & \colhead{($\mathrm{mK\,km\,s^{-1}}$)} & \colhead{} & \colhead{}}  
    \startdata
    45463.424 & $1\rightarrow0$ & 3 & 0.0 & -5.8546 & 2.0 & 364 & $>20$ & clean transition  \\
    90926.026 & $2\rightarrow1$ & 5 & 2.0 & -4.8723 & 2.5 & 854 & $>20$ & clean transition \\
    136387.028 & $3\rightarrow2$ & 7 & 6.6 & -4.3141 & 2.1 & 796 & $>20$ & clean transition \\
    \enddata
\end{deluxetable*}

%tab H2CS
\begin{deluxetable*}{ccccccccl}[h]
    \tablecaption{Lines of H$_2$CS detected toward G+0.693 with I$_{\mathrm{peak}}>0.8 \mathrm{K}$ and their corresponding spectroscopic information and parameters derived from the LTE fit.}
    \tablecolumns{9}
    \tablewidth{0pt}
    \tablehead{
    \colhead{Rest frequency} & \colhead{QNs} & \colhead{g$_u$} & \colhead{E$_u$} & \colhead{$\mathrm{\log A_{ul}}$} & \colhead{rms}& \colhead{$I_{peak}$} & \colhead{SNR} & \colhead{Comments} \\
    \colhead{(MHz)} & \colhead{} & \colhead{} & \colhead{(K)} & \colhead{(s$^{-1}$)} & \colhead{(mK)} & \colhead{(K)} & \colhead{} & \colhead{}}  
    \startdata
    101477.8048 & $3_{1,3}\rightarrow2_{1,2}$ & 21 & 2.0 & -4.8996 & 2.9 & 1.213 & $>20$ & clean transition \\
    103040.447 & $3_{0,3}\rightarrow2_{0,2}$ & 7 & 5.0 & -4.8285 & 2.6 & 0.951 & $>20$ & clean transition \\
    104617.027 & $3_{1,2}\rightarrow2_{1,1}$ & 27 & 18.2 & -4.8599 & 2.1 & 1.169 & $>20$ & clean transition  \\
    135298.26 & $4_{1,4}\rightarrow3_{1,3}$ & 27 & 22.9 & -4.4859 & 7.0 & 1.213 & $>20$ & clean transition \\
    137371.21 & $4_{0,4}\rightarrow3_{0,3}$ & 9 & 9.9 & -4.438  & 3.1 & 0.823 & $>20$ & clean transition \\
    139483.68 & $4_{1,3}\rightarrow3_{1,2}$ & 27 & 23.2 & -4.4462 & 2.7 & 1.197 & $>20$ & clean transition \\
    \enddata
\end{deluxetable*}

%table H2CO-18
\begin{deluxetable*}{ccccccccl}[h]
    \tablecaption{Lines of H$_2$C\isotope[18]{O} detected toward G+0.693 with their corresponding spectroscopic information and parameters derived from the LTE fit.}
    \tablecolumns{9}
    \tablewidth{0pt}
    \tablehead{
    \colhead{Rest frequency} & \colhead{QNs} & \colhead{g$_u$} & \colhead{E$_u$} & \colhead{$\mathrm{\log A_{ul}}$} & \colhead{rms}& \colhead{$\int T_A^* d\nu$} & \colhead{SNR} & \colhead{Comments} \\
    \colhead{(MHz)} & \colhead{} & \colhead{} & \colhead{(K)} & \colhead{(s$^{-1}$)} & \colhead{(mK)} & \colhead{($\mathrm{mK\,km\,s^{-1}}$)} & \colhead{} & \colhead{}}  
    \startdata
    134435.920 & $2_{1,2}\rightarrow1_{1,1}$ & 15 & 2.0 & -4.3361 & 6.7 & 839 & $>20$ & slightly blended with CH$_3$CHO \\
    138770.86 & $2_{0,2}\rightarrow1_{0,1}$ & 5 & 3.3 & -4.1697 & 2.7 & 814 & $>20$ & clean transition \\
    143213.07 & $2_{1,1}\rightarrow1_{1,0}$ & 15 & 15.3 & -4.2536 & 2.1 & 824 & $>20$ & clean transition  \\
    \enddata
\end{deluxetable*}

%fig ch318oh
\begin{figure*}[h]
    \centering
    \includegraphics[width=1\textwidth]{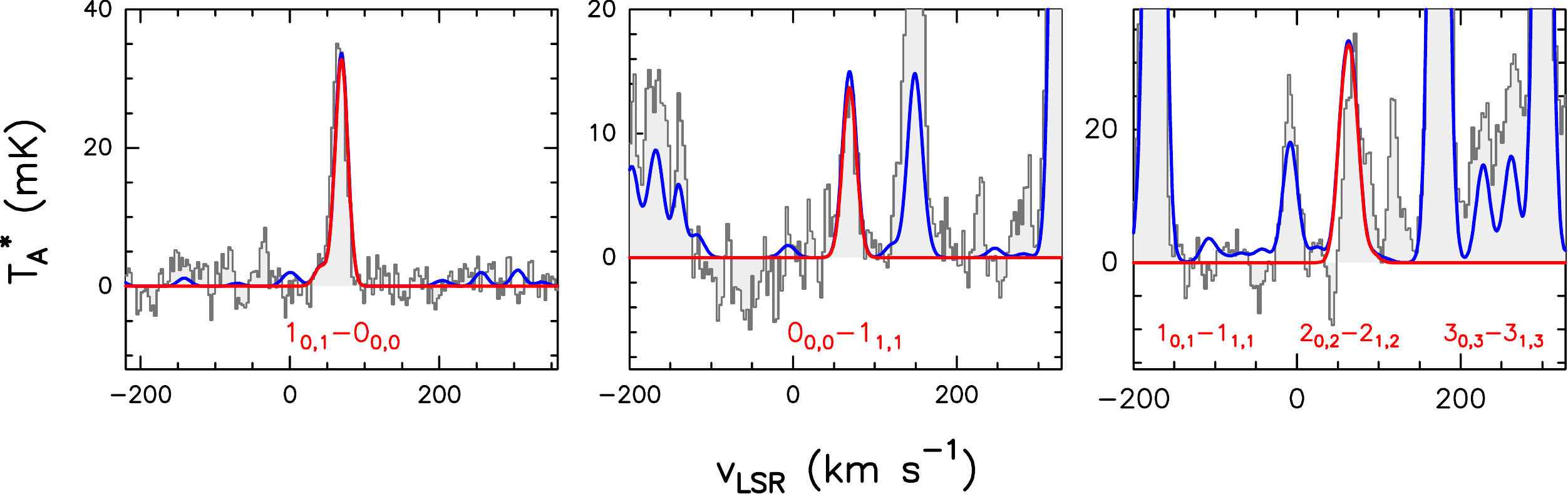}
    \caption{Selected lines of CH$_3$\isotope[18]{O}H detected toward G+0.693. Black, red and blue lines represent the observational data, the individual lines of CH$_3$\isotope[18]{O}H and the global fit considering all the species detected, respectively. See Table \ref{tab:ch3oh} for a complete description of their quantum numbers.}
    \label{fig:ch3oh}
\end{figure*}

%tab ch318oh
\begin{deluxetable*}{cccccccccl}[h]
    \tablecaption{Lines of CH$_3$\isotope[18]{O}H detected toward G+0.693 with their corresponding spectroscopic information and parameters derived from the LTE fit.\label{tab:ch3oh}}
    \tablecolumns{10}
    \tablewidth{0pt}
    \tablehead{
    \colhead{Rest frequency} & \colhead{QNs} & {T$_\mathrm{s}$} & \colhead{g$_u$} & \colhead{E$_u$} & \colhead{$\mathrm{\log A_{ul}}$} & \colhead{rms}& \colhead{$\int T_A^* d\nu$} & \colhead{SNR} & \colhead{Comments} \\
    \colhead{(MHz)} & \colhead{} & \colhead{} & \colhead{} &\colhead{(K)} & \colhead{(s$^{-1}$)} & \colhead{(mK)} & \colhead{($\mathrm{mK\,km\,s^{-1}}$)} & \colhead{} & \colhead{}}  
    \startdata
    46364.313 & $1_{0,1}\rightarrow0_{0,0}$ & A & 12 & 0.0 & -4.8985 & 2.8 & 709 & $>20$ & clean transition \\
    93505.902 & $2_{1,1}\rightarrow1_{1,0}$ & A & 20 & 16.6 & -5.6353 & 1.9 & 68 & 7 & blended with t-HC(O)SH\\
%    104329.446 & $0_{0,0}\rightarrow1_{1,1}$ & E & 4 & 7.8 & -4.8985 & 3.0 & 298 & 18 & clean transition  \\
    150698.06 & $1_{0,1}\rightarrow1_{1,1}$ & E & 12 & 7.8 & -4.7223 & 3.1 & 466 & $>20$ & clean transition \\
    150704.121 & $3_{0,3}\rightarrow3_{1,3}$ & E & 28 & 19.0 & -4.7330  & 3.1 & 127 & 8 & clean transition \\
    150704.173 & $2_{0,2}\rightarrow2_{1,2}$ & E & 20 & 12.3 & -4.7266 & 3.1 & 328 & 19 & clean transition \\
  %  164751.587 & $1_{1,0}\rightarrow1_{0,1}$ & E & 12 & 15.1 & -4.6413 & 4.2 & 107 & 18 & blended with CH$_3$COOH \\
  %  164756.128 & $2_{1,1}\rightarrow2_{0,2}$ & E & 20 & 19.5 & -4.6423 & 4.2 & 76 & 10 & blended with CH$_3$COOH \\
     \enddata
\end{deluxetable*}

%fig 13ch3ch2oh
\begin{figure*}[h]
    \centering
    \includegraphics[width=1\textwidth]{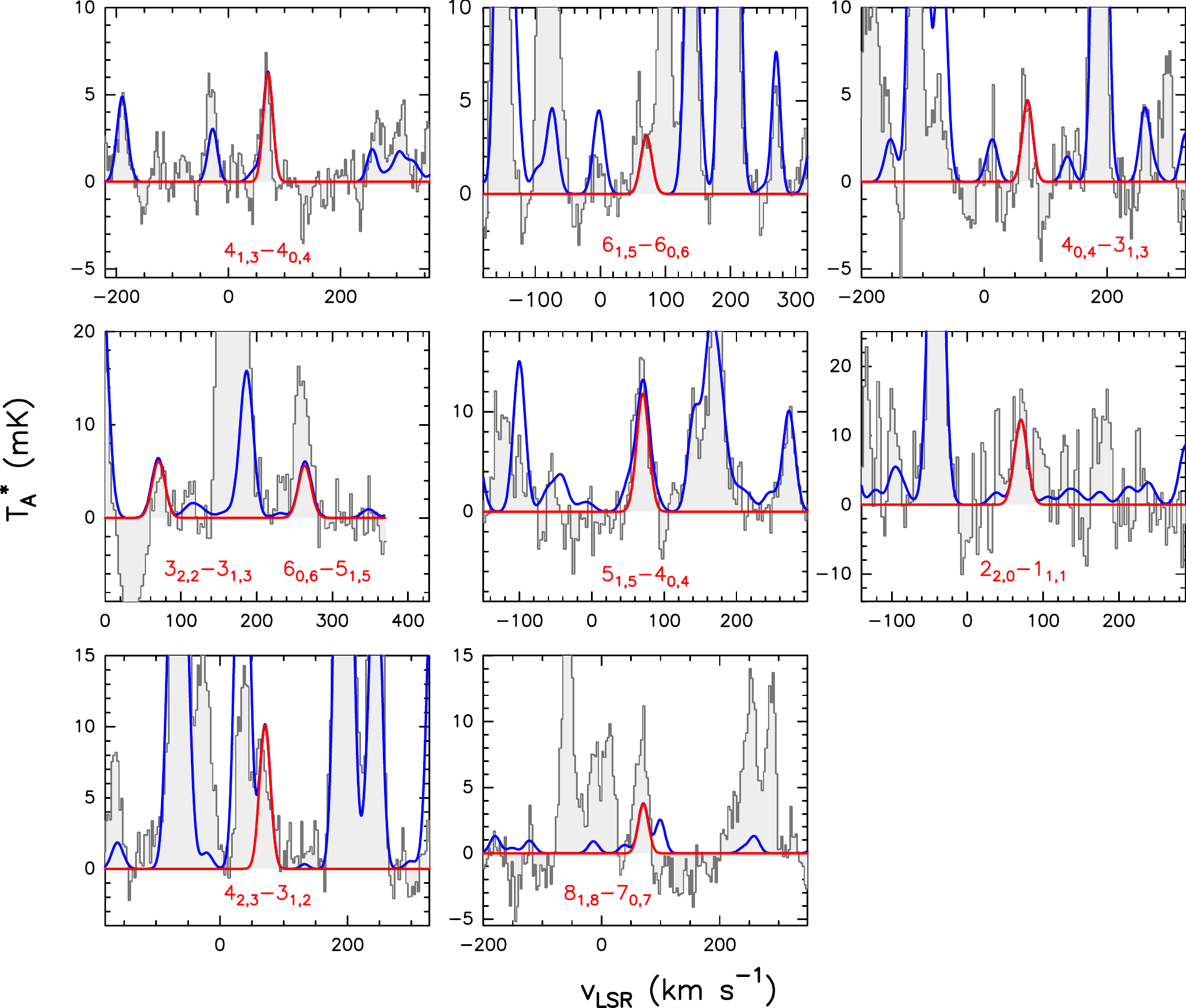}
    \caption{Selected lines of {\bf anti-}\isotope[13]{C}H$_3$CH$_2$OH detected toward G+0.693. Black, red and blue lines represent the observational data, the individual lines of \isotope[13]{C}H$_3$CH$_2$OH and the global fit with all species, respectively.}
    \label{fig:13ch3ch2oh}
\end{figure*}

%tab 13ch3ch2oh
\begin{deluxetable*}{ccccccccl}[h]
    \tablecaption{Lines of {\bf anti-}\isotope[13]{C}H$_3$CH$_2$OH detected toward G+0.693 with their corresponding spectroscopic information and parameters derived from the LTE fit. \label{tab:ch3ch2oh}}
    \tablecolumns{9}
    \tablewidth{0pt}
    \tablehead{
    \colhead{Rest frequency} & \colhead{QNs} & \colhead{g$_u$} & \colhead{E$_u$} & \colhead{$\mathrm{\log A_{ul}}$} & \colhead{rms}& \colhead{$\int T_A^* d\nu$} & \colhead{SNR} & \colhead{Comments} \\
    \colhead{(MHz)} & \colhead{} & \colhead{} & \colhead{(K)} & \colhead{(s$^{-1}$)} & \colhead{(mK)} & \colhead{($\mathrm{mK\,km\,s^{-1}}$)} & \colhead{} & \colhead{}}  
    \startdata
    32509.32 & $4_{1,3}\rightarrow4_{0,4}$ & 9 & 8.2 & -6.4293 & 1.3 & 134 & 19 & clean transition \\
    40435.51 & $6_{1,5}\rightarrow6_{0,6}$ & 13 & 17.0 & -6.2050 & 1.5 & 67 & 8 & slightly blended with unidentified transition \\
    44748.194 & $4_{0,4}\rightarrow3_{1,3}$ & 9 & 6.0 & -6.3778 & 2.1 & 100 & 9 & clean transition  \\
    82170.297 & $6_{0,6}\rightarrow5_{1,5}$ & 13 & 13.1 & -5.4772 & 3.5 & 121 & 6 & blended with unidentified transition \\
    82223.219 & $3_{2,2}\rightarrow3_{1,3}$ & 7 & 6.0 & -5.5620  & 2.8 & 130 & 9 & clean transition \\
    102934.258 & $5_{1,5}\rightarrow4_{0,4}$ & 11 & 8.2 & -5.1178 & 2.6 & 253 & 18 & clean transition \\
    113369.368 & $2_{2,0}\rightarrow1_{1,1}$ & 5 & 2.1 & -4.9878 & 6.8 & 262 & 7 & clean transition \\
    143297.176 & $4_{2,3}\rightarrow3_{1,2}$ & 9 & 6.3 & -4.8286 & 2.2 & 216 & 18 & clean transition \\
    144465.114 & $8_{1,8}\rightarrow7_{0,7}$ & 17 & 22.7 & -4.6339 & 1.7 & 81 & 9 & blended with unidentified transition \\
    \enddata
\end{deluxetable*}


\begin{thebibliography}{}

%Author(s), year, journal, volume, pages (aascitation)

\bibitem[Adriaens et al.(2010)]{adriens2010} Adriaens D.A., Goumans T.P.M., Catlow C.R.A, et al. \ 2010, J. Phys. Chem. C, 114, 1892.
\bibitem[Armijos-Abendaño et al.(2015)]{armijos2014} Armijos-Abendaño J., Mart{\'\i}n-Pintado, J., Requena-Torres M.A. et al.\ 2015, \mnras, 446, 3842
\bibitem[Asplund et al.(2009)]{asplund2009} Asplund M., Grevesse N., Sauval A. et al.\ 2009, \araa, 47 
%\bibitem[Bonfand et al.(2019)]{bonfand2019} Bonfand, M., Belloche, A., Garrod, R. T., et al. \ 2019, \aap, 628, A27.
%\bibitem[Cabezas et al.(2020)]{cabezas2020} Cabezas, C., Berm\'{u}dez, C., Tercero, B. et al.\ 2020, \aap, 639, A129
\bibitem[Calmonte et al.(2016)]{calmonte2016} Calmonte, U., Altwegg K., Balsiger H. et al.\ 2016, \mnras, 462, S253
\bibitem[Chandru et al.(2016)]{chandru2016} Chandru K., Gilbert A., Butch C., et al.\ 2016, Sci. Rep., 6, 29883
%\bibitem[Durig et al.(1975)]{durig1975raman} Durig J., Bucy W., Wurrey C., et al.\ 1975, J. Phys. Chem., 79, 988
\bibitem[Endres et al.(2016)]{CDMS} Endres C.P., Schlemmer S., Schilke P., et al.\ 2016, J. Mol. Spectrosc., 327, 95
%\bibitem[Feng et al.(2015)]{feng2015} Feng S., Beuther H., Henning Th., et al. \ 2015, \aap, 581, A71
\bibitem[Foden et al.(2020)]{foden2020prebiotic} Foden C.S., Islam S., Fern\'{a}ndez-Garc\'{\i}a C., et al.\ 2020, Science, 370, 865
\bibitem[Garrod \& Herbst(2006)]{garrod2006} Garrod R.T. and Herbst E. \ 2006, \aap, 457, 927
\bibitem[Gibb et al.(2000)]{gibb2000chemistry} Gibb E., Nummelin A., Irvine W.M., et al.\ 2000, \apj, 545, 309
\bibitem[Gorai et al.(2017)]{gorai2017} Gorai P., Das A., Sivaraman B., et al.\ 2017, \apj, 836, 70
\bibitem[Herbst \& Van Dishoeck(2009)]{herbst2009} Herbst E. and E.F. Van Dishoeck, \ 2009, \araa, 47, 427
\bibitem[Hocking et al.(1976)]{hocking1976rotational} Hocking W.H. and Winnewisser G., 1976, Z. Naturforsch, 31a, 995
\bibitem[Ioppolo et al.(2011)]{ioppolo2011} Ioppolo S., Van Boheemen Y., Cuppen H., et al.\ 2011, \mnras, 413, 2281
\bibitem[Jabri et al.(2020)]{jabri2020} Jabri A., Tercero B., Margul{\`e}s, L., et al. \ 2020, \aap, 644, A102
\bibitem[Jim\'{e}nez-Escobar et al.(2014)]{jm2011} Jim\'{e}nez-Escobar, A., Muñoz Caro, G.M. and  Chen, Y.J. \ 2014, \mnras, 443, 343
\bibitem[Jim\'{e}nez-Serra et al.(2020)]{ijimenez2020} Jim\'{e}nez-Serra I., Mart{\'\i}n-Pintado J., Rivilla V.M., et al.\ 2020, Astrobiology, 20, 1048
\bibitem[Kolesnikov\'{a} et al.(2014)]{kolesnikova2014} Kolesnikov\'{a} L., Tercero B., Cernicharo J. et al.\ 2014, \apjl, 784, L7
\bibitem[Laas et al.(2019)]{laas2019} Laas J.C. and Caselli P., 2019, \aap,624, A108
\bibitem[Lamberts(2018)]{lamberts2018} Lamberts T. 2018, \aap, 615, L2
\bibitem[Leman et al.(2017)]{leman2017} Leman L.J. and Ghadiri M.R., 2017, Synlett, 28, 68
%\bibitem[Lattelais et al.(2009)]{lattelais2009} Lattelais M., Pauzat F., Ellinger Y., et al. \ 2009, \apj, 696, L133
\bibitem[Linke et al.(1979)]{linke1979} Linke R., Frerking M.A. and Thaddeus P., 1979, \apj, 234, L139
\bibitem[Lohrmann et al.(1968)]{lohrmann1968} Lohrmann R. and Orgel L.E., 1968, Science, 161, 64
\bibitem[Majumdar et al.(2016)]{majumdar2016} Majumdar L., Gratier P., Vidal T., et al.\ 2016, \mnras, 458, 1859
\bibitem[Margul\`{e}s et al.(2020)]{margules2020} Margul\`{e}s L, Ilyushin V.V., McGuire B.A., et al. \ 2020, J. Mol. Spectrosc., 371, 111304
\bibitem[Mart\'{\i}n et al.(2019)]{madcuba} Mart\'{\i}n S., Mart\'{\i}n-Pintado J., Blanco-S\'{a}nchez, C. et al.\ 2019, \aap, 631, A159
\bibitem[Mart\'{\i}n et al.(2008)]{martin2008} Mart\'{\i}n S., Requena-Torres M.A., Mart{\'\i}n-Pintado J., et al.\ 2008, \apj, 678, 245
\bibitem[Motiyenko et al.(2020)]{motiyenko2020} Motiyenko R., Belloche A., Garrod R., et al.\ 2020, \aap, 642, A29
\bibitem[Muchowska et al.(2020)]{muchowska2020} Muchowska K.B. \& Moran J. \ 2020, Science, 370, 767-768
\bibitem[M\"{u}ller et al.(2016)]{muller2016} M\"{u}ller H.S.P., Belloche A., Li-Hong X., et al. 2016, \aap, 587, A92
\bibitem[Palumbo et al.(1997)]{palumbo1997} Palumbo M.E., T.R. Geballe and A.G.G.M. Tielens \ 1997, \apj, 479, 839
\bibitem[Parker et al.(2011)]{parker2011} Parker E.T., Cleaves H.J., Callahan M.P., et al.\ 2011, Orig. Life Evol. Biosph., 41, 201
\bibitem[Requena-Torres et al.(2006)]{requenatorres2006} Requena-Torres M.A., Mart{\'\i}n-Pintado J., Rodr{\'\i}guez-Fern{\'a}ndez N.J. et al.\ 2006, \aap, 455, 971
\bibitem[Requena-Torres et al.(2008)]{requena2008galactic} Requena-Torres M.A., Mart{\'\i}n-Pintado J., Mart{\'\i}n S. et al.\ 2008, \apj, 672, 352
\bibitem[Rivilla et al.(2019)]{rivilla2019_cyanomehtanimine} Rivilla V.M., Mart{\'\i}n-Pintado J., Jim{\'e}nez-Serra I., et al.\ 2019, \mnras, 483, L114
\bibitem[Rivilla et al.(2020)]{rivilla2020prebiotic} Rivilla V.M., Mart{\'\i}n-Pintado J., Jim{\'e}nez-Serra I., et al.\ 2020, \apjl, 899, L28
\bibitem[Rivilla et al.(2018)]{rivilla2018} Rivilla V.M., Jim{\'e}nez-Serra I., Zeng S., et al. \ 2018, \mnras, 475, L30 
\bibitem[Shalayel et al.(2020)]{shalayel2020} Shalayel I., Youssef-Saliba S., Vazart F., et al.\ 2020, Eur. J. Org. Chem, 20, 3019
\bibitem[Shingledecker et al.(2020)]{shingledecker2020sulphur} Shingledecker C.N., Lamberts T., Laas J.C. et al.\ 2020, \apj ,888, 52
\bibitem[Sinclair et al.(1973)]{sinclair1973} Sinclair, M.W., Fourikis, N., Ribes, J.C. et al. \ 1973, Aust. J. Phys., 26, 85
\bibitem[Tercero et al.(2021)]{tercero2020} Tercero F., L\'{o}pez-P\'{e}rez J.A., Gallego J.D., et al.\ 2021, \aap, 645, 22 
\bibitem[Tingle et al.(1991)]{tingle1991} Tingle N., Becker, C.H. and Ripudaman, M. \ 1991, Meteorit. Planet. Sci., 26, 117
\bibitem[Vidal et al.(2017)]{vidal2017} T.H.G. Vidal, J.-.C. Loison, A.Y. Jaziri., et al. \ 2017, \mnras, 469, 435
\bibitem[Wilson et al.(1999)]{wilson1999} Wilson T.L.\ 1999, Rep. Prog. Phys., 62, 143
\bibitem[Xu et al.(2012)]{xu2012} Xu L.H., Lees R.M., Grabbe G.T., et al.\ 2012, \jcp, 137, 104313
\bibitem[Zakharenko et al.(2019)]{zakharenko2019} Zakharenko O., Ilyushin V.V., Lewen F., et al. \ 2019, \aap, 629, A73
\bibitem[Zeng et al.(2018)]{zeng2018complex} Zeng S., Jim{\'e}nez-Serra I., Rivilla, V.M., et al.\ 2018, \mnras, 478, 2962
\bibitem[Zeng et al.(2020)]{zeng2020} Zeng S., Zhang Q., Jim{\'e}nez-Serra I., et al.\ 2020, \mnras, 497, 4896

\end{thebibliography}
\end{document}